\documentclass[12pt]{article}
\usepackage{putex}
\usepackage{graphicx}
\begin{document}
\preprint{PUPT-2470}

\title{Strings and vortex rings}
\authors{Steven S. Gubser, Revant Nayar, and Sarthak Parikh}
\institution{PU}{Joseph Henry Laboratories, Princeton University, Princeton, NJ 08544, USA}
\abstract{
We treat string propagation and interaction in the presence of a background Neveu-Schwarz three-form field strength, suitable for describing vortex rings in a superfluid or low-viscosity normal fluid.  A circular vortex ring exhibits instabilities which have been recognized for many years, but whose precise boundaries we determine for the first time analytically in the small core limit.  Two circular vortices colliding head-on exhibit stronger instabilities which cause splitting into many small vortices at late times.  We provide an approximate analytic treatment of these instabilities and show that the most unstable wavelength is parametrically larger than a dynamically generated length scale which in many hydrodynamic systems is close to the cutoff.  We also summarize how the string construction we discuss can be derived from the Gross-Pitaevskii lagrangian, and also how it compares to the action for giant gravitons.
}
\date{August 2014}

\maketitle

\tableofcontents

\section{Introduction}
\label{INTRODUCTION}

We are interested in the dynamics of vortex rings in a medium, moving slowly relative to the speed of sound $c_s$ and interacting with themselves through perturbations of the medium.  We will use the following action to describe the interacting vortices:
 \eqn{Sall}{
  S &= \sum_\alpha \left[ -c_s \tau_{1,\rm bare} 
      \int_{\Sigma_\alpha} dt \, d\theta \, |\partial_\theta \vec{X}_\alpha| + 
     \mu_1 \int_{\Sigma_\alpha} B_2 \right] 
     - {\lambda \over 2} \sum_{\alpha,\beta}
     \int_{\rm reg} dt \, d\theta \, d\tilde\theta {\partial_\theta \vec{X}_\alpha \cdot 
       \partial_{\tilde\theta} \vec{X}_\beta \over |\vec{X}_\alpha(\theta) - 
         \vec{X}_\beta(\tilde\theta)|} \,,
 }
where $\alpha$ and $\beta$ are labels running over the several separate vortices and $B_2$ is the pull-back of a spacetime gauge potential $B_2$ satisfying
 \eqn{BtwoExplicit}{
  dB_2 = H_3 = {\rho_0 \over 6} \epsilon_{mnp} dx^m \wedge dx^n \wedge dx^p \,.
 }
Essentially this action (but without the explicit tension term) was justified in \cite{Endlich:2013dma} as an effective description of hydrodynamical vortices.  Related actions were considered in the early literature on string theory, for example \cite{Nielsen:1973cs,Kalb:1974yc,Lund:1976ze}.  The tension term in \eno{BtwoExplicit} is understood to represent microscopic dynamics of the vortex core over which we do not have full control.  The regularized integral $\int_{\rm reg}$ provides some ultraviolet cutoff for the divergence that arises when the denominator of the integrand vanishes.  A common choice of regulator, which we will adopt, is to replace
 \eqn{Regularization}{
  |\vec{X}_\alpha(\theta) - \vec{X}_\beta(\tilde\theta)| \to 
    \sqrt{(\vec{X}_\alpha(\theta) - \vec{X}_\beta(\tilde\theta))^2 + a^2} \,,
 }
where the cutoff $a$ is approximately the radius of the vortex core.  We will assume $\mu_1>0$, which corresponds to a choice of orientation of the vortex; and it can be shown in the process of deriving \eno{Sall} that $\lambda>0$.

The action \eno{Sall} can be derived as the quasi-static approximation of classical effective string dynamics, where the effective strings move in response to a strong spatial background $H_3$ and interact with themselves through the exchange of electrical components $B_2$.  This classical effective string dynamics can in turn be derived from the Gross-Pitaevskii equation, under some simplifying assumptions and approximations.  There is in addition a weak coupling to a radiation field which can be represented as a perturbation $b_{ij}$ of $B_2$ and which propagates at the speed of sound.

Dynamics similar to \eno{Sall} have been studied for over a hundred years.  A notable early work is \cite{Dyson1893}, and modern reviews include \cite{Shariff92,Saffman92,Lim95,Donnelly09}.  We will start in section~\ref{SINGLE} by reviewing the instability of circular vortices \cite{Widnall73}.  We also calculate the zero point energy of fluctuations around circular vortices when it is well defined.  We will continue in section~\ref{COLLIDE} by treating the stronger instabilities that arise in head-on collisions of circular vortices \cite{Lim:1992fk}.  In both analyses we restrict ourselves to the limit of vanishingly small core size, so that we do not need to consider deformations of the core.  Such deformations are believed to play an important role in quantitatively accurate descriptions of both single vortex instabilities \cite{Saffman74,Widnall77,Saffman78} and the head-on collisions \cite{Lim95} in hydrodynamical settings.  A novelty of our treatment is that in the small core limit we achieve full analytical control over both the unperturbed solutions and their linearized perturbations in terms of elliptic integrals.

The relation between vortices and classical strings has received significant attention in the string theory and cosmology literature.  Early works \cite{Nielsen:1973cs,Kalb:1974yc,Lund:1976ze} emphasized the possible relevance to superfluid Helium, proposed a cosmological role for vortex defects (cosmic strings) in theories with broken global $U(1)$ symmetry, uncovered the role of the Neveu-Schwarz field $B_2$, and arrived at essentially the dynamics \eno{Sall}, including the tension term and a renormalization of it due to the regulated interaction term.  Derivations of the dynamics \eno{Sall} from effective theories of superfluids can be found in \cite{Davis89,Zee:1994qw}; see also \cite{Witten85} and the later work \cite{Franz:2006gb}.  For the sake of completeness, we will review in section~\ref{GROSS} a derivation of \eno{Sall} from the Gross-Pitaevskii action.  We then conclude in section~\ref{DISCUSSION} with a summary of results and a comparison of vortex ring phenomena to giant gravitons.  An appendix is devoted to a detailed comparison of single vortex results to an earlier study \cite{Widnall73}.

\section{Instabilities of a single circular vortex ring}
\label{SINGLE}

Let's parametrize a nearly circular vortex ring as follows:
 \eqn{TwoShapes}{
  \vec{X}(t,\theta) &= 
   \begin{pmatrix} (r(t) + \epsilon r_m(t) \cos m\theta) \cos\theta \\
    (r(t) + \epsilon r_m(t) \cos m\theta) \sin\theta \\
    z(t) + \epsilon z_m(t) \cos m\theta \end{pmatrix} \,.
 }
At $O(\epsilon^0)$ we will find a family of stationary solutions with constant $r$ and constant $\dot{z}$.  Next we will want to study linearized perturbations.  Plugging a perturbed ansatz like \eno{TwoShapes} into the action to obtain equations of motion is not necessarily justified, because in general the perturbations included in the ansatz may couple to others which are not included.  In this case, it is obvious from the axial symmetry of the unperturbed solution that perturbations with different $m$ cannot mix.  It is a matter of calculation to show that the perturbations shown in \eno{TwoShapes} do not mix with perturbations proportional to $\sin m\theta$.  We leave the details to the reader and here simply assert that in order to obtain correct evolution equations for $r_m$ and $z_m$, it is enough to plug \eno{TwoShapes} into the action \eno{Sall} and expand through $O(\epsilon^2)$.

For explicit calculations, we find it useful to work in a gauge where the background two-form gauge potential is
 \eqn{BtwoGauge}{
  B_2 = {\rho_0 \over 2} (X^1 dX^2 - X^2 dX^1) \wedge dX^3 \,,
 }
and to introduce a scaled lagrangian $L_{\rm one\ vortex}$ through the relation
 \eqn{Lscaled}{
  S = 2\pi \rho_0 \mu_1 \int dt \, L_{\rm one\ vortex} \,.
 }
Also we set
 \eqn{EtaLambda}{
  \eta_{\rm bare} = {c_s \tau_{1,\rm bare} \over \rho_0 \mu_1} \qquad\qquad
  \tilde\lambda = {\lambda \over \rho_0 \mu_1} \,.
 }
Through $O(\epsilon^2)$ we find
 \eqn{Lsplit}{
  L_{\rm one\ vortex} = L_0 + \epsilon^2 L_2 + \tilde\lambda r Q_0(q) + \epsilon^2 
    {\tilde\lambda \over 2r} \left[ Q_{rr}(q) r_m^2 + Q_{zz}(q) z_m^2 \right]
 }
where
 \eqn{Lfree}{
  L_0 &= -\eta_{\rm bare} r - {1 \over 2} r^2 \dot{z}  \cr
  L_2 &= -{\eta_{\rm bare} m^2 \over 4r} (r_m^2 + z_m^2) - 
    {1 \over 4} \dot{z} r_m^2 - {1 \over 2} r r_m \dot{z}_m
 }
and the remaining terms come from the interaction term in \eno{Sall} and depend on the dimensionless ratio
 \eqn{qDef}{
  q = {a \over r} \,.
 }
To determine $Q_0(q)$, we compare the general action \eno{Sall} to the desired form \eno{Lsplit} and arrive at
 \eqn{QzeroForm}{
  Q_0(q) = -{1 \over 4\pi r} \int d\theta \, d\tilde\theta \, 
    \left[ {\partial_\theta X^i(\theta) \partial_{\tilde\theta} X^i(\tilde\theta) \over
      \sqrt{a^2 + (\vec{X}(\theta) - \vec{X}(\tilde\theta))^2}} \right]_{O(\epsilon^0)} 
      \,.
 }
We are suppressing $t$ dependence for now, and we use $[A]_{O(\epsilon^n)}$ to denote the coefficient of $\epsilon^n$ in $A$.  To evaluate $Q_0(q)$, we work out the integral explicitly:
 \eqn{QzeroExplicit}{
  Q_0(q) &= -{1 \over 4\pi} \int d\theta \, d\tilde\theta \, 
    {\cos(\tilde\theta-\theta) \over \sqrt{q^2 + 2 - 2\cos(\tilde\theta-\theta)}}  \cr
   &= -{1 \over 4\pi} \int_0^{2\pi} d\theta \int_0^{2\pi} d\alpha 
    {\cos\alpha \over \sqrt{q^2 + 2 - 2\cos\alpha}}  \cr
   &= -{1 \over 2} \int_0^{2\pi} d\alpha {\cos\alpha \over \sqrt{q^2 + 2 - 2\cos\alpha}}
      \cr
   &= \int_{-1}^{1} {du \over v} \, {(u^2-v^2) \over
    \sqrt{q^2 +4u^2}}  \cr
   &= q E\left( -{4 \over q^2} \right) - \left( q + {2 \over q} \right)
       K\left( -{4 \over q^2} \right) \,.
 }
In the second line of \eno{QzeroExplicit} we set $\alpha = \tilde\theta - \theta$.  To get the third line we performed the $\theta$ integral (which is trivial in this case).  To get the fourth line we defined
 \eqn{uvDef}{
  \alpha = 2 \sin^{-1} u   \qquad\qquad v = \sqrt{1-u^2} \,.
 }
The fifth line of \eno{QzeroExplicit} involves the complete elliptic integrals $E$ and $K$.

The $O(\epsilon^2)$ terms are more involved but similar in concept.  By comparing \eno{Sall} and \eno{Lsplit}, we first extract
 \eqn{SecondCompare}{
  {1 \over 2} Q_{rr}(q) r_m^2 + {1 \over 2} Q_{zz}(q) z_m^2 = 
    -{r \over 4\pi} \int d\theta \, d\tilde\theta \,
    \left[ {\partial_\theta X^i(\theta) \partial_{\tilde\theta} X^i(\tilde\theta) \over
      \sqrt{a^2 + (\vec{X}(\theta) - \vec{X}(\tilde\theta))^2}} \right]_{O(\epsilon^2)} 
      \,.
 }
Using the same sequence of operations exhibited in \eno{QzeroExplicit}, we next obtain
 \eqn{Lint}{
  Q_{rr}(q) &=\int_{-1}^{1} {du \over v} \left[ 
    {n_{5/2} \over (q^2+4u^2)^{5/2}} + 
    {n_{3/2} \over (q^2+4u^2)^{3/2}} + 
    {n_{1/2} \over \sqrt{q^2+4u^2}}
   \right]  \cr
  Q_{zz}(q) &=  \int_{-1}^{1} {du \over v} \left[ 
    -{2(u^2-v^2) u^2 U_{m-1}^2(v) \over (q^2+4u^2)^{3/2}} - 
    {m^2 T_{2m}(v) \over \sqrt{q^2+4u^2}} \right] \,,
 }
where $T_n(v)$ and $U_n(v)$ are Chebyshev polynomials of the first and second kind, respectively, and
 \eqn{ThreeNs}{
  n_{5/2} &= 24u^4(u^2-v^2)T_{m}^2(v) \cr
  n_{3/2} &=  1+2u^2-8u^4+(-1+8u^2-12u^4)T_{2m}(v)-8mu^4v\,U_{2m-1}(v) \cr
  n_{1/2} &= (1+m^2)(u^2-v^2)T_{2m}(v)+4mu^2v\,U_{2m-1}(v) \,.
 }
The integrals in \eno{Lint} can be expressed in terms of the same elliptic functions that enter into \eno{QzeroExplicit}, but for subsequent calculations we will only need the small $q$ expansions:
 \eqn{Qexpand}{
  Q_0(q) &= \log {qe \over 8} + 1 + O(q^2 \log q)  \cr
  Q_{rr}(q) &= \frac{m^2}{2} \log {qe \over 8} + \frac{1}{8}R_{rr} + O(q^2 \log q)  \cr
  Q_{zz}(q) &= \frac{m^2}{2} \log {qe \over 8} + \frac{1}{8}R_{zz} + O(q^2 \log q) \,,
 }
where
 \eqn{Rdefs}{
  R_{zz} &= (4m^2-3) \tilde{H}_{m-1/2} - 2m^2  \cr
  R_{rr} &= (4m^2-1) \tilde{H}_{m-1/2} -2(m^2+2)
 }
and
 \eqn{HarmonicNumbers}{
  \tilde{H}_n \equiv \psi(n+1)-\psi(1) + 2\log 2 \qquad\hbox{and}\qquad
   \psi(n) = \Gamma'(n)/\Gamma(n) \,.
 }
For odd half-integer $n$, $\tilde{H}_n$ is a rational number.

The $\log q$ terms in \eno{Qexpand} are divergent in the $a \to 0$ limit.  This UV divergence comes from interactions of segments of a vortex ring which are very close to one other.  The divergence can be cured by regarding the tension term in \eno{Sall} as a counterterm.  To be precise, after dropping terms which vanish in the $q \to 0$ limit, all remaining dependence on $a$ and the rescaled tension $\eta_{\rm bare}$ comes from dependence on the length scale
 \eqn{ellZeroDef}{
  \ell_0 \equiv {a \over 8} e^{1-\eta_{\rm bare}/\tilde\lambda} \,.
 }
Thus, at least formally, we may take a limit in which $a \to 0$ and $\eta_{\rm bare} \to -\infty$ with $\ell_0$ held fixed, and in this limit, higher order terms in $q$ (for instance, the ones indicated as $O(q^2 \log q)$ in \eno{Qexpand}) vanish.  The existence of such a limit is appealing from the standpoint of effective field theory, because it indicates that \eno{Sall} is renormalizable with only the tension counterterm.  It is a somewhat peculiar limit from the standpoint of hydrodynamics; as we will review later, in some standard hydrodynamical contexts, $a$ and $\ell_0$ are separated by a factor of order unity, not some large hierarchy.  In order to use the small $q$ expansion \eno{Qexpand}, we only need $r \gg a$.  This is certainly implied if we work in the small $a$, fixed $\ell_0$ limit, with $r/\ell_0$ also held fixed.

It is straightforward to see that the equations of motion for the unperturbed vortex rings ($\epsilon=0$) are
 \eqn{UnperturbedEoms}{
  \dot{r} = 0 \qquad\qquad \dot{z} = -{\tilde\lambda \over r} \log {r \over \ell_0} \,,
 }
and that the linearized equations of motion for perturbations are
 \eqn{LinearizedEoms}{
  \dot{r}_m - {\tilde\lambda z_m \over 4r^2} 
   \left[ 4m^2 \log {r \over \ell_0} - R_{zz} \right] &= 0  \cr
  \dot{z}_m + {\tilde\lambda r_m \over 4r^2}
   \left[ 4(m^2-1) \log {r \over \ell_0} - R_{rr} \right] &= 0 \,.
 }
Thus $r_m(t)$ and $z_m(t)$ undergo harmonic motion with frequency
 \eqn{omegaE}{
  \omega_m = {\tilde\lambda \over 4r^2} \sqrt{
    \left[ 4m^2 \log {r \over \ell_0} - R_{zz} \right]
    \left[ 4(m^2-1) \log {r \over \ell_0} - R_{rr}  \right]
   } \,.
 }
There is an instability if $\omega_m$ is imaginary.  Noting that $R_{zz}=0$ for $m=0$, we find $\omega_0 = 0$; and noting that $R_{rr}=0$ for $m=1$, we find $\omega_1 = 0$.  These results are expected since the $m=0$ and $1$ modes constitute infinitesimal shifts among unperturbed solutions.

Essentially the results \eno{omegaE} were obtained in \cite{Widnall73} through more traditional hydrodynamical methods, though the exact result \eno{omegaE} was not obtained.  In Appendix~\ref{COMPARE} we compare the approximate treatment of \cite{Widnall73} with \eno{omegaE} and find good agreement.  Here let us note the main qualitative feature: the $m$-th mode is unstable when the radius is in some finite range of values close to
 \eqn{rSatisfies}{
  r \approx 4m \ell_0 e^{\gamma-1/2} \approx 4.321 m \ell_0 \,,
 }
where $\gamma \approx 0.5772$ is the Euler-Mascheroni constant.  The range of values of $r$ over which the $m$-th mode is unstable is broad when $m$ is small---so broad that the $m=2$ and $m=3$ instabilities overlap.  At larger $m$, the instabilities become progressively narrower.  Another way of writing \eno{rSatisfies} is that the reduced wavelength of unstable modes (when they exist) is
 \eqn{LambdaSatisfies}{
  {\lambda \over 2\pi} = {r \over m} \approx 4 e^{\gamma-1/2} \ell_0 \approx 4.321 \ell_0 \,.
 }
The fact that $\lambda/2\pi$ is significantly greater than $\ell_0$ might suggest that our treatment doesn't depend entirely on having a large separation of scales $a \ll \ell_0$.  However, the results of \cite{Widnall73} were immediately criticized in \cite{Saffman74} on grounds that in realistic hydrodynamic settings, the wavelength \eno{LambdaSatisfies} is insufficiently large to justify the vortex filament approach.  Eventually a more detailed, finite-core-size analysis appeared \cite{Widnall77} which features good agreement with experiment.  Our approach in this paper is to focus on the regime of small $a$ and fixed $\ell_0$, even though it is not immediately applicable in hydrodynamic settings.

\subsection{Zero point energy of string fluctuations}

Classically, the energy and momentum in the $z$ direction of a circular vortex ring are $\epsilon = 2\pi\rho_0\mu_1 \tilde\epsilon$ and $p = 2\pi\rho_0\mu_1 \tilde{p}$, where
 \eqn{EandP}{
  \tilde\epsilon = \tilde\lambda r \left( \log {r \over \ell_0} - 1 \right) \qquad\qquad
  \tilde{p} = -{r^2 \over 2} \,.
 }
The Hamiltonian relation $\dot{z} = \partial\epsilon / \partial p$ \cite{Donnelly70} is easily checked.  In a quantum mechanical setting, it is interesting to inquire what the contribution to the energy is from zero point energy (analogous to the Luscher term for the QCD string \cite{Luscher:1980ac}, see also \cite{Alvarez:1981kc}), and with the explicit frequencies \eno{omegaE} in hand, we can answer this question.  There are two oscillation modes for each value of $m$, one corresponding to the $\cos m\theta$ perturbations indicated in \eno{TwoShapes}, and the other corresponding to replacing $\cos m\theta \to \sin m\theta$.  The zero point energy of these modes is
 \eqn{ZPEdef}{
  {\rm Z.P.E.} = \hbar \sum_{m=2}^\infty \omega_m \,.
 }
(The sum may be extended to include $m=1$ and/or $m=0$ since $\omega_m=0$ for these modes.)  In fact, the sum should be cut off in some way to reflect the fact that modes with wavelength shorter than $a$ should not be included.  A suitable regulator is to include a factor $e^{-mq}$ inside the sum, where $q=a/r$ as in \eno{qDef}.  Note that this regulator disappears in the limit $a \to 0$ with $\ell_0$ held fixed, while in this limit the $\omega_m$ remain constant.\footnote{If \eno{Sall} is augmented to include a full relativistic Nambu term (or similar inertial effects without full relativistic invariance), then terms quadratic in $\dot{r}_m$ and $\dot{z}_m$ will enter into $L_2$, but suppressed by $1/c_s^2$.  With such terms present, the equations of motion become second order, and oscillation modes break into two sets, fast and slow.  The slow modes are the ones whose frequencies $\omega_m$ we have found, and they are scarcely disturbed by the introduction of inertial effects.  The fast modes have frequencies scaling as $c_s^2$.  In computing the zero point energy \eno{ZPEdef}, we are including the contribution of the slow modes only.}

\begin{figure}
  \centerline{\includegraphics[width=5in]{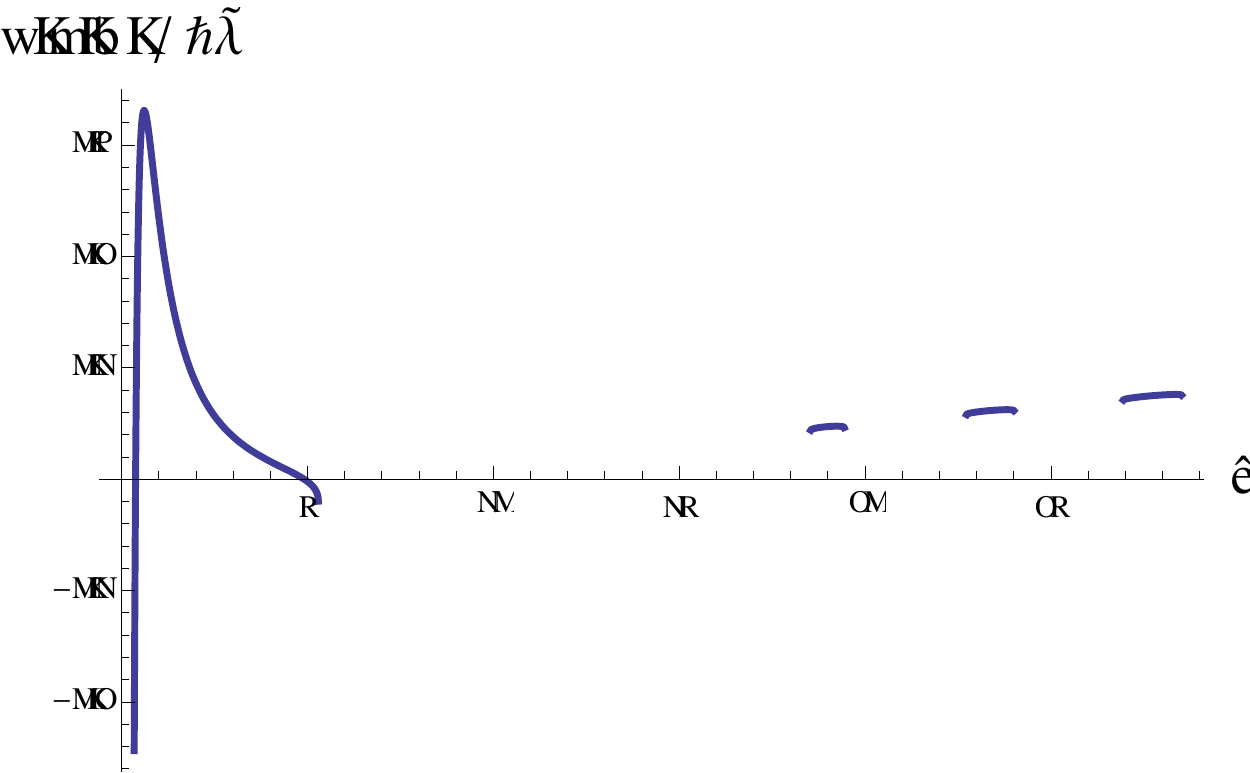}}
  \caption{The zero point energy for a circular vortex loop in zeta function regularization.  Gaps in the plot are regions of instability.}\label{ZPEzeta}
 \end{figure}

Besides ultraviolet divergences, another problem with \eno{ZPEdef} is the complexity of the summand, which as far as we can see precludes an analytic expression for the sum.  To proceed, we split
 \eqn{omegaSplit}{
  \omega_m = {\tilde\lambda \over r^2} (\Omega^{\rm series}_m + \Omega^{\rm remainder}_m)
 }
where
 \eqn{omegaSeries}{
  \Omega^{\rm series}_m &= \left( m^2 - {1 \over 2} \right) \left( 
    \log m - \log {r \over \ell_0} \right) + 
    \left( 2 \log 2 + \gamma - {1 \over 2} \right) m^2  \cr
   &\qquad\qquad{} 
   - {1 \over 2} \log m - \log 2 - {\gamma \over 2} - {11 \over 24} \,.
 }
The reason for the specific choice \eno{omegaSeries} is that $\sum_m \Omega^{\rm remainder}_m$ is then absolutely convergent, with no regulator required, and so it may be computed through direct numerical summation.  Using zeta function regularization, in particular the formulas
 \eqn{ZetaFormulas}{
  \sum_{m=1}^\infty 1 &= \zeta(0) = -{1 \over 2}  \cr
  \sum_{m=1}^\infty m^2 &= \zeta(-2) = 0  \cr
  \sum_{m=1}^\infty \log m &= -\zeta'(0) = {1 \over 2} \log 2\pi  \cr
  \sum_{m=1}^\infty m^2 \log m &= -\zeta'(-2) = {\zeta(3) \over 4\pi^2} \,,
 }
we obtain the result
 \eqn{OmegaSeriesSum}{
  {\rm Z}^{\rm series}_{\rm zeta} \equiv 
   \sum_{m=1}^\infty \Omega^{\rm series}_m = 
     -{1 \over 4} \log {\pi r \over 2\ell_0} + 
      {\zeta(3) \over 4\pi^2} + {\gamma \over 4} + {11 \over 48} \,.
 }
Combining this result with a numerical summation of the convergent terms leads to the total zero point energy plotted in figure~\ref{ZPEzeta}.

\section{Head-on collisions of two circular vortex rings}
\label{COLLIDE}

Let us now consider the collision of two vortex rings.  The first one will be parametrized just as in \eno{TwoShapes}.  We will restrict attention here to a second ring whose shape is the mirror image of the first:
 \eqn{MirrorShape}{
  \tilde{\vec{X}}(t,\tilde\theta) &= 
   \begin{pmatrix} (r(t) + \epsilon r_m(t) \cos m\tilde\theta) \cos\tilde\theta \\
    (r(t) + \epsilon r_m(t) \cos m\tilde\theta) \sin\tilde\theta \\
    -z(t) - \epsilon z_m(t) \cos m\tilde\theta \end{pmatrix} \,.
 }
It is convenient for $\tilde\theta$ to wind around the second vortex ring in the same direction that $\theta$ winds around the first; however, physically we have in mind the opposite orientation for the second ring, so that with negative $\dot{z}$ the rings will approach one another over time.  Therefore, when plugging \eno{TwoShapes} and \eno{MirrorShape} into the action \eno{Sall}, we have to insert an extra minus sign on the $\int B_2$ term for the second string, and also on the term describing interactions between the strings.  Altogether, $S = 2\pi\kappa\mu_1 \int dt \, L_{\rm two\ vortices}$ where
 \eqn{FullLagCollide}{
  L_{\rm two\ vortices} &= 2 L_{\rm one\ vortex} - 
     2 \tilde\lambda r Q_0\left(\sqrt{q^2+s^2}\right)  \cr
   &\qquad{} + \epsilon^2 {\tilde\lambda \over 2r}
    \left[ S_{rr}(s,q) r_m^2 + 2S_{rz}(s,q) r_m z_m + 
      S_{zz}(s,q) z_m^2 \right] \,.
 }
On the right hand side of  $L_{\rm two\ vortices}$ is the lagrangian defined in \eno{Lsplit}, and the additional terms come from the interaction of the strings with each other.  We have introduced a new dimensionless variable,
 \eqn{sDef}{
  s = {2z \over r} \,,
 }
and we have defined $S_{rr}$, $S_{rz}$, and $S_{zz}$ through 
 \eqn{SecondCompareS}{
  {1 \over 2} S_{rr}(s,q) r_m^2 + S_{rz}&(s,q) r_m z_m + 
    {1 \over 2} S_{zz}(s,q) z_m^2  \cr
   &= 
    {r \over 2\pi} \int d\theta \, d\tilde\theta \,
    \left[ {\partial_\theta X^i(\theta) 
      \partial_{\tilde\theta} \tilde{X}^i(\tilde\theta) \over
      \sqrt{a^2 + (\vec{X}(\theta) - 
        \tilde{\vec{X}}(\tilde\theta))^2}} \right]_{O(\epsilon^2)} 
      \,.
 }
Through the same process that we showed in \eno{QzeroExplicit}, we find
\eqn{SIntegrals}{
S_{rr}(s,q) &= -2\, Q_{rr}\left(\sqrt{q^2+s^2}\right)  \cr
S_{rz}(s,q) &= s\int_{-1}^1 du\, {2\over v}\left[
			-{12 u^2(u^2-v^2)T_m^2(v) \over(q^2+s^2+4u^2)^{5/2}}
			+ {2(u^2-v^2)T_m^2(v)+2mu^2v\, U_{2m-1}(v) \over (q^2+s^2+4u^2)^{3/2}} 
			\right]  \cr
S_{zz}(s,q) &=  \int_{-1}^{1} du\, {2\over v}\left[
			-{6 s^2 (u^2-v^2) T_m^2(v) \over (q^2+s^2+4u^2)^{5/2}} 
			+ {2(u^2-v^2)T_m^2(v) \over (q^2+s^2+4u^2)^{3/2}} 
			- {m^2 T_{2m}(v) \over \sqrt{q^2+s^2+4u^2}} \right] \,,
} 
where $u$ and $v$ are defined in \eno{uvDef}.  We will show below that the unperturbed colliding vortices stay a distance $\sim \ell_0$ away from one another.  Therefore, if we work in the now-familiar limit $a \to 0$ with $\ell_0$ held fixed, we may set $q=0$ and write the functions $S_{ij}$ appearing in \eno{SIntegrals} in terms of $s$ only:
 \eqn{EllipticForms}{
  S_{ij}(s) = S_{ij}^{(E)}(m;s) E\left(-{4 \over s^2}\right) + 
    S_{ij}^{(K)}(m;s) K\left(-{4 \over s^2}\right) \,,
 }
where $S_{ij}^{(E)}(m;s)$ and $S_{ij}^{(K)}(m;s)$ are rational functions of $s$ with integer coefficients depending on $m$.  We do not have a general formula for the $S_{ij}^{(E)}(m;s)$ and $S_{ij}^{(K)}(m;s)$, but they may be worked out straightforwardly from \eno{SIntegrals} for any given value of $m$.

Let's first use \eno{FullLagCollide} to analyze the motion of unperturbed colliding vortex rings.  The equations of motion following from $L_{\rm two\ vortices}$ at $O(\epsilon^0)$ are
 \eqn{eomsZero}{
  r^2 \dot{s} + 2 \tilde\lambda \log {r \over \ell_0} + 2 \tilde\lambda Q_0(s) &= 0  \cr
  r \dot{r} - 2 \tilde\lambda Q_0'(s) &= 0 \,,
 }
where dots denote derivatives with respect to $t$, while primes denote derivatives with respect to $s$.  If we use $s$ in place of $t$ as the independent variable, then \eno{eomsZero} simplifies to
 \eqn{rsZero}{
  {dr \over ds} = {-r \, Q_0'(s) \over Q_0(s) + \log {r \over \ell_0}} \,.
 }
This differential equation may be solved as
 \eqn{rSoln}{
  r(s) = {\xi_0 e \ell_0 \over W_0(\xi_0 e^{Q_0(s)})} \qquad\hbox{where}\qquad
    \xi_0 = {r_0 \over e\ell_0} \log {r_0 \over e\ell_0} \,.
 }
(We will assume $r_0 > e\ell_0$.)  Here $W_0(y)$ is the principal branch of the Lambert $W$ function, which is defined implicitly through the equation
 \eqn{LambertWDef}{
  y = W e^W \,.
 }
The principal branch $W_0(y)$ is positive for positive $y$.  Note that we did not use any special properties of $Q_0(s)$ to derive \eno{rSoln}: mild smoothness assumptions suffice for the derivation of \eno{rSoln}.  The parameter $r_0$ in \eno{rSoln} is the initial radius at early times, when the vortices are far apart.  Starting from \eno{rSoln} it is straightforward to show that at late times, when $s$ is small and $r$ is large, we have
 \eqn{LateTimeZ}{
  z \to z_{\rm min} \equiv {4\ell_0 \over e} \qquad\hbox{from above,}
 }
and also
 \eqn{dotRValue}{
  \dot{r} \approx {\tilde\lambda \over z} \approx 
     {e \over 4} {\tilde\lambda \over \ell_0} \,.
 }
The result \eno{rSoln} is not new; in fact, a similar problem was solved in \cite{Dyson1893}, and related work appears in \cite{Shariff89}.

We will now use the perturbations in \eno{TwoShapes} to study the late time stability of the solution we just found for head-on collisions of circular vortex rings.  Starting from the $O(\epsilon^2)$ terms of \eno{FullLagCollide}, and using the zeroth order equations of motion \eno{eomsZero}, it is straightforward to derive linearized equations which take the form
 \eqn{PerturbsForm}{
  -s {d \over ds} \begin{pmatrix} r_m \\ z_m \end{pmatrix} = 
   w \begin{pmatrix} r_m \\ z_m \end{pmatrix} \,,
 }
where $w$ is a $2 \times 2$ matrix whose form is slightly complicated.  To express it, we write
 \eqn{wMatDef}{
 w = w_1+w_2+w_3+w_4
 }
where
 \eqn{wvalues}{\seqalign{\span\TL & \span\TR & \qquad\qquad\span\TL & \span\TR}{
  {w_1 \over w_F} &= 
    m^2 \log {\ell_0 \over r} \begin{pmatrix} 0 & 1 \\ -1 & 0 \end{pmatrix}
  			+ {1\over 4}\begin{pmatrix} 0 & R_{zz} \\ -R_{rr} & 0 \end{pmatrix}  &
  {w_2 \over w_F} &= \log {\ell_0 \over r}
    \begin{pmatrix} 0 & 0 \\ 1 & 0 \end{pmatrix}  \cr
  {w_3 \over w_F} &= 
    \begin{pmatrix} 2Q'_0(s) & 0 \\ -Q_0(s) + s Q'_0(s) & 0 \end{pmatrix} &
  {w_4 \over w_F} &= \begin{pmatrix} S_{rz}(s) & S_{zz}(s) \\ 
    -S_{rr}(s) & -S_{rz}(s) \end{pmatrix}
 }}
and
 \eqn{wFDef}{
  w_F = -{1\over 2}\,{s \over Q_0(s)+\log{r\over \ell_0}} \,.
 }
Let $\Gamma$ be the eigenvalue of $w$ with the largest real part.  The corresponding mode varies with time as $s^{-\Gamma}$.  Noting that $s \to 0$ and $r \to \infty$ as $t \to \infty$, we see that a positive real part of $\Gamma$ corresponds to an instability.  For fixed $s$ and sufficiently large $m$, both eigenvalues of $w$ have real part equal to $-1/2$; thus short wavelength perturbations damp out.  But for small $s$, large $r$, and $m$ not too large, there are strong instabilities.  The purpose of the rest of this section is to give an approximate account of these instabilities.

In handling the small $s$, large $r$ limit, our first step is to use the expansion
 \eqn{SexpansionsQ}{
  Q_0(s) &= \log {se \over 8} + 1 + O(s^2 \log s)
 }
It is then straightforward to show that
 \eqn{GammaSmallS}{
  \Gamma \approx 
    \Gamma_1 \equiv {1 \over 1 + \log {z \over z_{\rm min}}} 
      \left( -{1 \over 2} + {1 \over 2} \sqrt\Delta \right)
 }
where
 \eqn{DeltaDef}{
  \Delta &\equiv (1 + s S_{rz}(s))^2  \cr
   &\qquad{} - {s^2 \over 16} 
   \Big[ 4m^2 \log {r \over \ell_0} - R_{zz} - 4 S_{zz}(s) \Big]
   \left[ 4m^2 \log {r \over \ell_0} - R_{rr} - 4 S_{rr}(s) - 
      4 \log {z \over z_{\rm min}} \right] \,.
 }
To go further, we need approximate forms for the $S_{ij}$.  We were unable to find approximations which work uniformly well for all $m$; instead, we found the expansions
\eqn{Sexpansions}{
	S_{rr}(s) &= -m^2 \log {s \over 8} -\left(m^2-{1\over 4}\right) \tilde{H}_{m-1/2} -{m^2\over 2}+1-{3\over 8}\left(m^2-{1\over 2}\right)(m^2-2)s^2 \log {s\over 8} \cr
	& \qquad{} -{3 s^2\over 8}\left(m^2-{1\over 4}\right)\left(m^2-{9\over 4}\right)\tilde{H}_{m-1/2}+{s^2\over 64}\left(22m^4-7m^2-44\right)+ O(s^4 \log s) \cr
	S_{rz}(s) &= -\frac{2}{s}  -\frac{3}{2} \left( m^2-\frac{1}{2}\right)s\log {s\over 8}- {3s\over 2}\left(m^2-{1 \over 4}\right) \tilde{H}_{m-1/2} \cr
	& \qquad{} +{s\over 4}\left(m^2+{11\over 2}\right) +O\left(s^3\log s\right)\cr
		S_{zz}(s) &= {4\over s^2}+\left(m^2-\frac{3}{2}\right)\log{s\over 8} + \left(m^2-{3  \over 4}\right)\tilde{H}_{m-1/2} - {m^2 \over 2}-{11\over 4} \cr
	& \qquad{} 	+{1\over 8}\left(m^4-{23\over 2}m^2+{45\over 8}\right)s^2\log {s\over 8}+{s^2\over 8}\left(m^2-{1\over 4}\right)\left(m^2-{45\over 4}\right)\tilde{H}_{m-1/2} \cr
	& \qquad{} +{s^2\over 64}\left(-10m^4+33m^2+{291\over 4}\right)+O\left(s^4\log s\right)
}
which are valid at small $s$ and fixed $m$ but may fail when the product $ms$ is not small.  Plugging these approximations into $\Gamma_1$ gives an explicit but complicated estimate for the growth rate which we will refer to as $\Gamma_2$.

\begin{figure}
  \centerline{\includegraphics[width=5in]{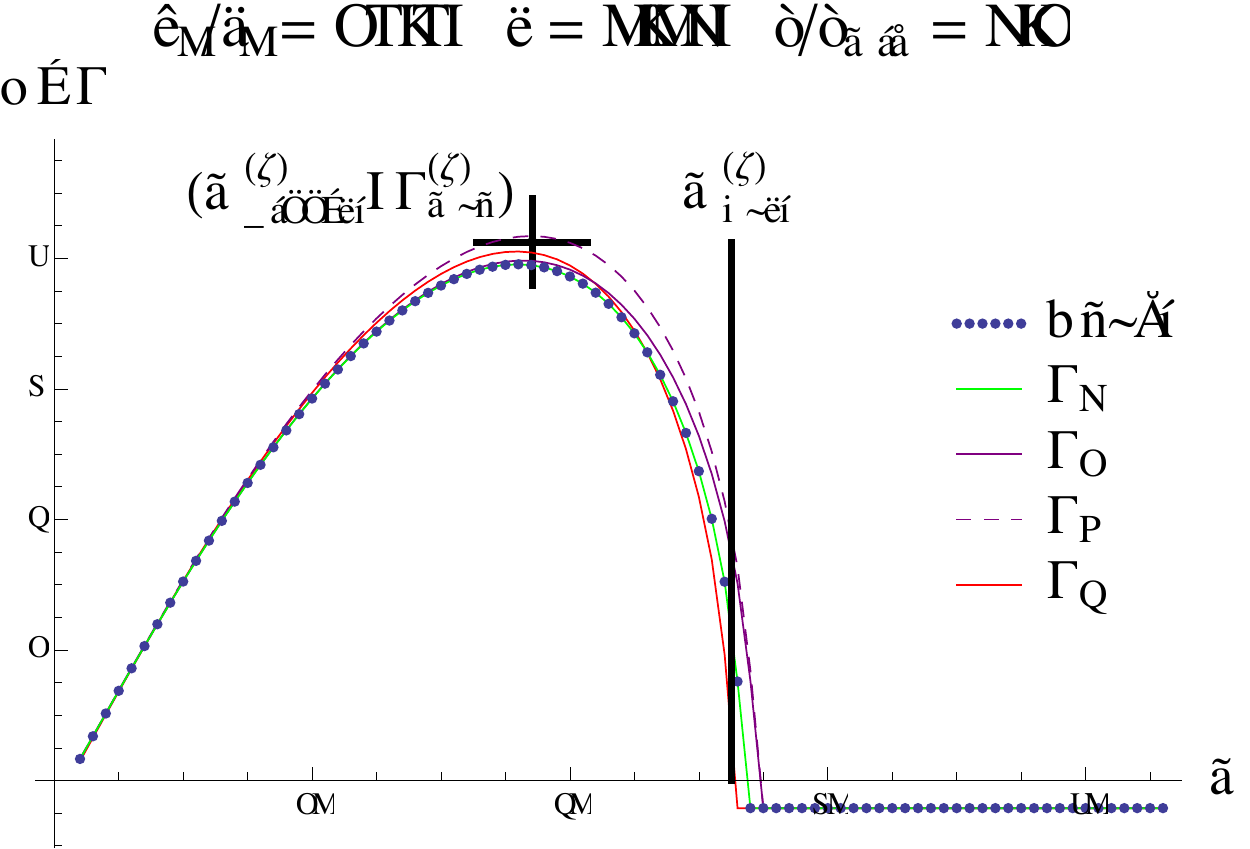}}
  \caption{(Color online.)  Instabilities of colliding vortex rings.  Each point above the horizontal axis represents an unstable mode, while the points below the axis represent stable modes.  The estimates $\Gamma_1$ through $\Gamma_4$ are explained in the main text, from \eno{GammaSmallS} to \eno{GammaFour}.  Formulas for the estimates $m_{\rm Biggest}^{(\zeta)}$ and $m_{\rm Last}^{(\zeta)}$ of the most unstable and last unstable mode are given in \eno{mBiggestSmallz}, and the estimate $\Gamma_{\rm max}^{(\zeta)}$ of the maximum growth rate is given in \eno{GammaMax}.  The black lines intersecting at $(m_{\rm Biggest}^{(\zeta)},\Gamma_{\rm max}^{(\zeta)})$ are intended to guide the eye rather than to indicate error bars.}\label{ExampleInstability}
 \end{figure}

It turns out that the leading approximations $S_{rz}(s) \approx -2/s$ and $S_{zz} \approx 4/s^2$ are sufficient to understand the main aspects of the late time dynamics.  Using these leading approximations for $S_{rz}(s)$ and $S_{zz}(s)$, we arrive at the relatively simple formula
 \eqn{SrrForm}{
  \Delta \approx 4m^2 \log {r \over \ell_0} - R_{rr} - 4 S_{rr}(s) - 
    4 \log {z \over z_{\rm min}} + 1 \,.
 }
Plugging the approximate expression for $S_{rr}(s)$ in \eno{Sexpansions} into \eno{SrrForm} and then plugging the result into \eno{GammaSmallS} gives an estimate for $\Gamma$ which we will refer to as $\Gamma_3$.  Finally, for a more concise expression, we expand \eno{SrrForm} at large $m$ with $ms$ held fixed to extract\footnote{In fact, the expansion described in the main text results in $\Delta \approx 4m^2 \left( \log {z \over z_{\rm min}} + {3m^2 s^2 \over 8} \log {ms e^{\gamma-11/12} \over 2} \right)$.  The simpler expression \eno{SimpleForm} works approximately as well in practice.}
 \eqn{SimpleForm}{
  \Delta \approx \Delta_4 \equiv 4m^2 \left( \log {z \over z_{\rm min}} + 
     {3m^2 s^2 \over 8} \log {ms e^{\gamma-1} \over 2} \right) \,.
 }
Thus our final, simplest approximation to the growth rate is
 \eqn{GammaFour}{
  \Gamma \approx \Gamma_4 \equiv {1 \over 1 + \log {z \over z_{\rm min}}}
    \left( -{1 \over 2} + m \sqrt{\log {z \over z_{\rm min}} + 
     {3m^2 s^2 \over 8} \log {ms e^{\gamma-1} \over 2}} \right) \,.
 }
At late times, a range of modes is unstable, starting with $m=2$ and extending up to some fairly large value of $m$, as illustrated in figure~\ref{ExampleInstability}.
By solving $\Delta_4=0$ at fixed $s$ and $z$, we can extract an estimate of the last unstable mode: 
 \eqn{mLast}{
  m_{\rm Last} = {2 e^{1-\gamma} \over s} \exp\left\{ {1 \over 2} 
    W_{-1}\left( -{4 \over 3} e^{-2(1-\gamma)} \log {z \over z_{\rm min}} \right)
      \right\} \,;
 }
and by solving $d\Delta_4/dm = 0$ at fixed $s$ and $z$, we can extract an estimate of the most unstable mode:
 \eqn{mBiggest}{
  m_{\rm Biggest} = {2 e^{{3 \over 4}-\gamma} \over s} \exp\left\{ {1 \over 2} 
    W_{-1}\left( -{2 \over 3} e^{-2\left( {3 \over 4}-\gamma \right)} 
      \log {z \over z_{\rm min}} \right)
      \right\} \,.
 }
Here $W_{-1}(y)$ is the lower branch of the Lambert $W$ function, which takes values between $-1$ and $-\infty$ for $y \in (-1/e,0)$.  Thus far we have not required $z/z_{\rm min}$ parametrically close to $1$; however, in order for the $\Gamma_4$ estimate in \eno{GammaFour} to make sense, we should have 
 \eqn{zRatioRange}{
  1 < z/z_{\rm min} \lsim \exp\left\{ {3 \over 4} e^{1-2\gamma} \right\} \approx 1.9 \,,
 }
since for larger $z/z_{\rm min}$ the estimate \eno{mLast} breaks down on account of $W_{-1}$ becoming complex when its argument is less than $-1/e$. 
It is observed that the $\Gamma_1$ estimate in \eno{GammaSmallS} works fairly well over a much broader range, for $z/z_{\rm min}$ as large as $10$ and/or $s$ as large as $1/2$.\footnote{An interesting phenomenon captured correctly by $\Gamma_1$ at larger values of $\log {z \over z_{\rm min}}$ and $s$ is that there can be two disjoint regions of instability, one at small $m$ and another at $m$ comparable to the location of the single vortex instability as indicated in \eno{rSatisfies}-\eno{LambdaSatisfies}.  As time progresses, these two regions broaden and merge, and at late times one enters the regime well described by the $\Gamma_4$ estimate.}

If we add the assumption that $z/z_{\rm min}$ is close to $1$, then the argument of $W_{-1}$ is small, and we may approximate
 \eqn{WmOneApprox}{
  W_{-1}(y) \approx \log(-y) - \log(-\log(-y)) + 
    {\log(-\log(-y)) \over \log(-y)} \,.
 }
The last term in \eno{WmOneApprox}, as well as additional corrections to \eno{WmOneApprox}, vanish in the limit $y \to 0^-$.  Using \eno{WmOneApprox}, we find
 \eqn{mBiggestSmallz}{
  m_{\rm Biggest} &\approx m_{\rm Biggest}^{(\zeta)} \equiv
   {\zeta^{-{1+\zeta \over 2\zeta}} \over s} 
   \sqrt{{8 \over 3} \log {z \over z_{\rm min}}}  \cr 
  m_{\rm Last} &\approx m_{\rm Last}^{(\zeta)} \equiv 
    \sqrt{2}\, m_{\rm Biggest}^{(\zeta)} \,,
 }
where
 \eqn{zetaDef}{
  \zeta \equiv {3 \over 2} - 2\gamma + \log {3 \over 2} - 
    \log\left( \log {z \over z_{\rm min}} \right) \,.
 }
$\zeta$ is a positive number which slowly grows large as $z/z_{\rm min} \to 1$.
With a little more work, one can show that the maximum value of $\Gamma$ can be approximated as
 \eqn{GammaMax}{
  \Gamma_{\rm max} \approx \Gamma_{\rm max}^{(\zeta)} \equiv 
   {1 \over 1 + \log {z \over z_{\rm min}}}
    \left( -{1 \over 2} + {2 \over s} {\log {z \over z_{\rm min}} \over
     \sqrt{{3 \over 2} + 3 \zeta - 6 
      \log\left( \zeta^{-{1+\zeta \over 2\zeta}} \right)}} \right) \,.
 }
Our main qualitative conclusions follow from \eno{mBiggestSmallz} and \eno{GammaMax}:
 \begin{itemize}
  \item The most unstable mode at late times, when $z/z_{\rm min}$ is close to $1$ has reduced wavelength significantly greater than $\ell_0$.  More precisely:
 \eqn{BigWavelength}{
  {\lambda \over 2\pi} = {r \over m_{\rm Biggest}} \approx 
   {\sqrt{24} \over e}
    {\zeta^{1+\zeta \over 2\zeta} \over \sqrt{\log {z \over z_{\rm min}}}} \ell_0 \,.
 }
The trend of this expression as $z/z_{\rm min} \to 1$ from above is dominated by the $1/\sqrt{\log {z \over z_{\rm min}}}$; in other words, $\lambda / 2\pi \gg \ell_0$ by approximately this factor.  This is in contrast to the single vortex instability, where according to \eno{LambdaSatisfies} $\lambda / 2\pi > \ell_0$ by only an $O(1)$ factor.
  \item The most unstable mode can be very unstable, due to the $1/s$ dependence in $\Gamma_{\rm max}$.  The strongest $z$ dependence in $\Gamma_{\rm max}$ comes from its behaving approximately as ${2 \over s} \log {z \over z_{\rm min}}$.  For a fixed initial radius $r_0$, one can show starting from \eno{rSoln} that
 \eqn{sLogzLimit}{
  {2 \over s} \log {z \over z_{\rm min}} \approx {e \over 4} {r_0 \over \ell_0}
    \log {r_0 \over \ell_0 e} \qquad\hbox{at late times} \,.
 }
The instability is strongest, then, when $r_0 \gg \ell_0$.  It is almost completely absent for $r_0$ only slightly larger than $\ell_0 e$.  At very late times, the factor in \eno{GammaMax} written in terms of $\zeta$ suppresses the instability, even for large $r_0$; however this happens very slowly.  The example in figure~\ref{ExampleInstability} makes it clear that an unremarkable choice of parameters leads to $\Gamma_{\rm max}$ significantly larger than $1$.
 \end{itemize}

It should be noted that this late-time analysis controls the dominant instability only in a limit where the initial perturbations are sufficiently small.  Numerical exploration of \eno{PerturbsForm} indicates that perturbations with finite though small amplitude may grow large before one reaches the asymptotic regime with $z/z_{\rm min}$ close to $1$.

\section{Derivation of the effective string action}
\label{GROSS}

Although well-established (e.g.~in \cite{Davis89,Zee:1994qw,Franz:2006gb}, an effective description of the dynamics of superfluids in terms of effective strings coupled to a Neveu-Schwarz two-form gauge potential $B_2$ is crucial to our analysis, and it is therefore worth reviewing here, along the lines of \cite{Zee:1994qw}.

One standard starting point for describing the collective dynamics of a superfluid is the Gross-Pitaevskii lagrangian,
 \eqn{LGP}{
  {\cal L}_{\rm GP} = i \phi^\dagger \partial_t \phi - {1 \over 2m} (\nabla\phi)^2 -
    {g \over 2} \left( |\phi|^2 - \rho_0 \right)^2 \,.
 }
The $U(1)$ symmetry is spontaneously broken:
 \eqn{phiPolar}{
  \phi = \sqrt\rho e^{i\eta} \,,
 }
where $\rho \approx \rho_0$ and $\eta$ is a Goldstone boson.  Then
 \eqn{LGPPolar}{
  {\cal L}_{\rm GP} = {\cal L}_1 + {\cal L}_2
 }
where
 \eqn{LoneLtwo}{
  {\cal L}_1 = -\rho\dot\eta - {\rho \over 2m} (\nabla\eta)^2  \qquad\qquad
  {\cal L}_2 = -{\left(\nabla\rho\right)^2 \over 8m\rho} - {g \over 2} (\rho-\rho_0)^2 \,.
 }
The first step is to dualize $\eta$ to $B_2$.  For simplicity, we first treat \eno{LGPPolar} under the assumption that $\eta$ takes values in ${\bf R}$ rather than ${\bf R} / 2\pi {\bf Z}$: This amounts to omitting the strings.  Let us introduce a four-vector $f^\mu = (\rho,\vec{f})$, where $\vec{f}$ are auxiliary fields over which we integrate.  It is easily verified that
 \eqn{LOneMod}{
  {\cal L}_1 + {m \over 2\rho} \left( \vec{f} - {\rho \over m} \nabla\eta \right)^2 = 
    -f^\mu \partial_\mu \eta + {m \over 2\rho} \vec{f}^2 \,.
 }
Thus
 \eqn{JustXi}{
  e^{i\int d^4 x \, {\cal L}_1} &= 
   \int {\cal D} \vec{f} \, \exp\left\{i \int d^4 x \, \left[ {\cal L}_1 + 
    {m \over 2\rho} \left( \vec{f} - {\rho \over m} \nabla\eta \right)^2 \right] 
    \right\}  \cr
  &= \int {\cal D} \vec{f} \, \exp\left\{i \int d^4 x \, \left[ 
    -f^\mu \partial_\mu \eta + {m \over 2\rho} \vec{f}^2 \right] \right\} \,.
 }
where we have defined the measure and contour so that
 \eqn{GaussianMeasure}{
  \int {\cal D} \vec{f} \, \exp\left\{ i \int d^4 x \, {m \over 2\rho} \vec{f}^2 \right\}
    = 1 \,.
 }
The requirement \eno{GaussianMeasure} is troublesome because the result of the Gaussian integration depends on $\rho$, which is a fluctuating field.  We must assume that $\rho \approx \rho_0$ is a good enough approximation (away from vortices) that \eno{GaussianMeasure} can be enforced; in other words, additional terms from relaxing \eno{GaussianMeasure} to include $\rho$ dependence are ignored.

Our next step is to perform the path integral over $\eta$.  The first term on the right hand side of \eno{LOneMod} may be integrated by parts, so that the path integration over $\eta$ simply enforces the constraint 
 \eqn{fConstraint}{
  \partial_\mu f^\mu = 0 \,.
 }
We solve \eno{fConstraint} by setting
 \eqn{fDual}{
  f^\mu = {1 \over 6} \epsilon^{\mu\nu\lambda\sigma} H_{\nu\lambda\sigma}
 }
where $H_3 = dB_2$.  The reason that \eno{fDual} works is that \eno{fConstraint} is implied by the equality of mixed partials acting on $B_2$.  \eno{fDual} means in particular that
 \eqn{OneTerm}{
  {m \over 2\rho} \vec{f}^2 = {m \over 4\rho} \sum_{i,j} H_{0ij}^2 \,.
 }
This is the electric part of the standard action for $B_2$.  To get the rest of the action, we note that
 \eqn{LtwoH}{
  {\cal L}_2 = -{g \over 12} h_{ijk}^2 - {(\nabla h_{ijk})^2 \over 48 m\rho} \,,
 }
where we have split
 \eqn{Hsplit}{
  H_{\nu\lambda\sigma} = H^{(0)}_{\nu\lambda\sigma} + h_{\nu\lambda\sigma}
 }
and set
 \eqn{Hzero}{
  H^{(0)}_{123} = \rho_0
 }
with all other components of $H^{(0)}_{\nu\lambda\sigma}$ vanishing except the ones related to \eno{Hzero} by index permutation.  Our convention in \eno{LtwoH} and elsewhere is to sum over indices without restriction.  We can still set $H_3 = dB_2$, and we split $B_2 = B^{(0)}_2 + b_2$.  To summarize, $\rho_0$ is the background superfluid density, and $h_{ijk}$ describes density fluctuations around it, and (nearly) the whole claim is
 \eqn{WholeClaim}{
  \int {\cal D}\rho &{\cal D}\eta \, \exp\left\{i \int d^4 x \, \left[ -\rho\dot\eta - 
    {\rho \over 2m} (\nabla\eta)^2 - {(\nabla\rho)^2 \over 8m\rho} - 
      {g \over 2} (\rho-\rho_0)^2 \right] \right\}  \cr
   &= \int {\cal D} B_2 \, \exp\left\{i \int d^4 x \, \left[ 
     -{g \over 12} \eta^{\mu\alpha} \eta^{\nu\beta} \eta^{\lambda\gamma}
       h_{\mu\nu\lambda} h_{\alpha\beta\gamma} - {(\nabla h_{ijk})^2 \over 48 m\rho_0}
      \right] \right\} \,,
 }
where $\eta^{\mu\nu} = \diag\{ -{1 \over c_s^2},1,1,1 \}$ and
 \eqn{csGP}{
  c_s^2 = {g\rho_0 \over m} \,.
 }
We now drop the $(\nabla h_{ijk})^2$ term because it affects the dispersion relation only in the UV:
 \eqn{Dispersion}{
  \omega^2 = c_s^2 k^2 + {\# k^4 \over m^2} = c_s^2 k^2 
    \left( 1 + \# k^2 a_{\rm GP}^2 \right) \,,
 }
where $\#$ is a factor of order unity, and
 \eqn{aGP}{
  a_{\rm GP} \equiv {1 \over \sqrt{2mg\rho_0}}
 }
is the Gross-Pitaevskii coherence length, which is also the approximate radius of a vortex core.  In writing \eno{WholeClaim} we have on the right hand side replaced $\rho \to \rho_0$ so that the final theory of $B_2$ excitations is free.  Interactions with variable $\rho$ could be developed perturbatively.  Note also that we are assuming that the Jacobian between ${\cal D}\rho {\cal D}\eta$ and the natural measure ${\cal D} B_2$ for the two-form gauge field can be neglected.  Of course, to properly describe the path integral over $B_2$, one must develop some sort of gauge-fixing technology.

The main way in which \eno{WholeClaim} is incomplete is that we ignored the possibility of vortices, around which one has a winding of the phase, $\eta \to \eta + 2\pi$.  Restricting ourselves for simplicity to a single vortex whose spacetime embedding is $X^\mu(\tau,\sigma)$, one can show that
 \eqn{MixedPartials}{
  \epsilon^{\lambda\sigma\mu\nu} \partial_\mu \partial_\nu \eta = -2\pi
    \int d^2 \sigma \, \epsilon^{ab} \partial_a X^\lambda \partial_b X^\sigma
     \delta^4(x^\mu - X^\mu(\tau,\sigma)) \,.
 }
For any fixed string configuration $X^\mu(\tau,\sigma)$, we can split
 \eqn{etaSplit}{
  \eta = \eta_{\rm vortex} + \eta_{\rm smooth} \,,
 }
where $\eta_{\rm vortex}$ satisfies \eno{MixedPartials} and $\eta_{\rm smooth}$ is a smooth function (i.e.~satisfying $\epsilon^{\lambda\sigma\mu\nu}\partial_\mu \partial_\nu \eta_{\rm smooth} = 0$).  Then
 \eqn{fdEta}{
  -f^\mu \partial_\mu \eta = -f^\mu \partial_\mu \eta_{\rm vortex} - 
    f^\mu \partial_\mu \eta_{\rm smooth} \,.
 }
Inside the path integral, the second term in \eno{fdEta} can be integrated by parts, and $\eta_{\rm smooth}$ can be treated as the lagrangian multiplier enforcing the constraint $\partial_\mu f^\mu = 0$.  Upon doing this, we may employ \eno{fDual} to write
 \eqn{fdVortex}{
  -f^\mu \partial_\mu \eta_{\rm vortex} &= -{1 \over 2} \epsilon^{\mu\nu\lambda\sigma}
      \partial_\nu B_{\lambda\sigma} \partial_\mu \eta_{\rm vortex}
    = -{1 \over 2} B_{\lambda\sigma} \epsilon^{\lambda\sigma\mu\nu} 
      \partial_\mu \partial_\nu \eta_{\rm vortex}  \cr
   &= \pi \int d^2 \sigma \, \epsilon^{ab} \partial_a X^\lambda \partial_b X^\sigma
       B_{\lambda\sigma} \delta^4(x^\mu - X^\mu(\tau,\sigma)) \,,
 }
where in the second step we dropped a total derivative.  Finally, we integrate over ${\bf R}^{3,1}$:
 \eqn{fdEtaInt}{
  -\int d^4 x \, f^\mu \partial_\mu \eta_{\rm vortex} = 2\pi \int B_2 \,,
 }
which is the desired form, with
 \eqn{muValue}{
  \mu_1 = 2\pi \,.
 }

An additional aspect of vortex dynamics is that the core has some energy cost per unit length.  Without entering into detail, we parametrize this microscopic dynamics by the tension term appearing in \eno{Sall}.  Putting this tension term together with \eno{WholeClaim} and \eno{fdEtaInt}, we find the effective action
 \eqn{Seff}{
  S_{\rm eff} = \sum_\alpha \left[ -c_s \tau_{1,\rm bare} 
      \int_{\Sigma_\alpha} dt \, d\theta \, |\partial_\theta \vec{X}_\alpha| + 
     \mu_1 \int_{\Sigma_\alpha} B_2 \right] - 
   \int d^4 x \, {g \over 2} h_3^2 \,,
 }
where $h_3^2 = {1 \over 6} \eta^{\mu\alpha} \eta^{\nu\beta} \eta^{\lambda\gamma} h_{\mu\nu\lambda} h_{\alpha\beta\gamma}$.  Already to write the tension term we have assumed that the speed of all parts of the vortex is much less than $c_s$; otherwise we would need the full Nambu action or some generalization thereof.  In order to obtain the action \eno{Sall} which we use in actual calculations, we must integrate out $h_3$, using again the quasi-static approximation for the motion of the vortices.  An efficient means of doing so is to re-introduce the spacetime vortex current 
 \eqn{Jagain}{
  j^{\lambda\sigma}(x^\mu) = \sum_\alpha \mu_1 \int_{\Sigma_\alpha} d^2\sigma \, 
    \epsilon^{ab} 
    \partial_a X^\lambda_\alpha \partial_b X^\sigma_\alpha 
      \delta^4(x^\mu-X^\mu_\alpha(\tau,\sigma)) \,.
 }
This current already appeared in \eno{MixedPartials}.  Omitting the coupling to the background field $B_2^{(0)}$, the relevant terms in \eno{Seff} are 
 \eqn{Sj}{
  \sum_\alpha \mu_1 \int_{\Sigma_\alpha} b_2 - \int d^4 x \, {g \over 2} h_3^2 =
    \int d^4 x \, \left[ -{g \over 2} h_3^2 + {1 \over 2} b_{\mu\nu} j^{\mu\nu}
      \right] \,.
 }
To integrate out $h_3$ we must include some gauge-fixing terms.  A convenient choice for present purposes is to require
 \eqn{hThreeReplace}{
  \partial_i b_{i\mu} = 0
 }
where $i$ runs from $1$ to $3$ and $\mu$ runs from $0$ to $3$.  Recalling $h_{\mu\nu\lambda} = 3\partial_{[\mu} b_{\nu\lambda]}$, we see that
 \eqn{hThreeExpand}{
  h_3^2 &= -{1 \over 2c_s^2} (\partial_0 b_{ij} + \partial_j b_{0i} + 
     \partial_i b_{j0})^2 + 
    {1 \over 6} (\partial_i b_{jk} + \partial_j b_{ki} + \partial_k b_{ij})^2  \cr
   &= -{1 \over c_s^2} (\partial_j b_{0i})^2 - {1 \over 2c_s^2} (\partial_0 b_{ij})^2 + 
     {1 \over 2} (\partial_k b_{ij})^2 + \hbox{(total derivatives)}
 }
where the gauge condition \eno{hThreeReplace} has been used in the second equality.

For static strings, the only non-vanishing components of $j^{\mu\nu}$ are $j^{0i}$.  We therefore consider the combination
 \eqn{SCoulombic}{
  S_{\rm Coulombic} &\equiv \int d^4 x \, \left[ b_{0i} j^{0i} + 
       {g \over 2 c_s^2} (\partial_j b_{0i})^2 \right]  \cr
    &\hskip-0.5in{} = \int d^4 x \, \left[ -{g \over 2c_s^2} 
      \left( b_{0i} - {c_s^2 \over g} \square_{{\bf R}^3}^{-1} j^{0i} \right) 
        \square_{{\bf R}^3}
      \left( b_{0i} - {c_s^2 \over g} \square_{{\bf R}^3}^{-1} j^{0i} \right) + 
      {c_s^2 \over 2g} j^{0i} \square_{{\bf R}^3}^{-1} j^{0i} \right] \,,
 }
where $\square_{{\bf R}^3} \equiv \partial_j^2$ is the laplacian on ${\bf R}^3$.  Integrating out $b_{0i}$ enables us to drop the first term in square brackets in the last expression in \eno{SCoulombic}; or in the classical theory, we would set
 \eqn{bSet}{
  b_{0i}(t,\vec{x}) = 
    -{c_s^2 \over 4\pi g} \int d^3 y \, {j^{0i}(t,\vec{y}) \over |\vec{x}-\vec{y}|} \,,
 }
where we have noted that
 \eqn{SquareGreen}{
  \square_{{\bf R}^3} {-1 \over 4\pi |\vec{x}|} = \delta^3(x) \,.
 }
Thus we arrive at
 \eqn{SGreen}{
  S_{\rm Coulombic} &= -{c_s^2 \over 8\pi g} \int dt \, d^3 x \, d^3 y \, 
     {j^{0i}(t,\vec{x}) j^{0i}(t,\vec{y}) \over |\vec{x}-\vec{y}|}  \cr
   &= -{c_s^2 \mu_1^2 \over 8\pi g} \sum_{\alpha,\beta} \int dt \, 
     d\theta \, d\tilde\theta {\partial_\theta \vec{X}_\alpha \cdot 
       \partial_{\tilde\theta} \vec{X}_\beta \over |\vec{X}_\alpha(\theta) - 
         \vec{X}_\beta(\tilde\theta)|} \,,
 }
where to obtain the second line we have parametrized strings in static gauge, so that $X^0_\alpha = t$ and $\vec{X}_\alpha = \vec{X}_\alpha(t,\theta)$ with $\theta \in (0,2\pi)$; thus
 \eqn{CurrentOnceMore}{
  j^{0i}(t,\vec{x}) = \sum_\alpha \mu_1 \int d\theta \, 
    \partial_\theta X^i_\alpha \delta^3(\vec{x}-\vec{X}_\alpha(t,\theta)) \,.
 }
The above treatment is complete for truly static strings, where all time dependence is trivial, and the field $b_{ij}$ decouples because $j^{ij} = 0$.  For quasi-static strings, moving much slower than the speed $c_s$, $j^{ij}$ is a small but non-vanishing source for $b_{ij}$, and the correct treatment of \eno{Seff} is
 \eqn{SeffFinal}{
  S_{\rm eff} &= \sum_\alpha \left[ -c_s \tau_{1,\rm bare} 
      \int_{\Sigma_\alpha} dt \, d\theta \, |\partial_\theta \vec{X}_\alpha| + 
     \mu_1 \int_{\Sigma_\alpha} B^{(0)}_2 \right]  \cr
   &\qquad{} - 
   {c_s^2 \mu_1^2 \over 8\pi g} \sum_{\alpha,\beta} \int dt \, 
     d\theta \, d\tilde\theta {\partial_\theta \vec{X}_\alpha \cdot 
       \partial_{\tilde\theta} \vec{X}_\beta \over |\vec{X}_\alpha(\theta) - 
         \vec{X}_\beta(\tilde\theta)|}  \cr
   &\qquad{} + 
    \int d^4 x \left[ {1 \over 2} b_{ij} j^{ij} + 
      {g \over 4c_s^2} (\partial_0 b_{ij})^2 - {g \over 4} (\partial_k b_{ij})^2
      \right] \,.
 }
In other words, we have quasi-static strings interacting Coulombically with one another and coupled to a phonon field.  Returning to \eno{Sall}, we see that the match with the first two lines of \eno{SeffFinal} is precise once we identify $B_2$ in \eno{Sall} as the background field $B^{(0)}_2$ and set
 \eqn{MuLambdaValues}{
  \lambda = {c_s^2 \mu_1^2 \over 4\pi g} = \pi {c_s^2 \over g}
   = \pi {\rho_0 \over m} \,,
 }
where in the second equality we used \eno{muValue} and in the third we used \eno{csGP}.  Note that upon use of \eno{EtaLambda} we obtain
 \eqn{TildeLambda}{
  \tilde\lambda = {1 \over 2m} \,.
 }

As a consistency check, we can inquire whether the motion of vortices as determined through a classical treatment of \eno{Sall} is indeed much less than $c_s$.  Referring to \eno{UnperturbedEoms} and \eno{csGP}, we find
 \eqn{vRatio}{
  {v^2 \over c_s^2} = {a_{\rm GP}^2 \over 2r^2} \left( \log {r \over \ell_0} \right)^2
    \,,
 }
which is indeed much less than $1$ provided $r \gg a_{\rm GP}$.  On the other hand, \eno{dotRValue} shows that the expansion that follows head-on vortex collision proceeds at a speed comparable to
 \eqn{vtyp}{
  v_{\rm typ} = {\tilde\lambda \over \ell_0} \,.
 }
In a Gross-Pitaevskii treatment,
 \eqn{vtypRatio}{
  {v_{\rm typ}^2 \over c_s^2} = {1 \over 4 g \rho_0 m \ell_0^2} = 
    {a_{\rm GP}^2 \over 2\ell_0^2}
 }
is {\it not} small: indeed, the study \cite{Roberts71} shows that $a \approx a_{\rm GP}$ is moderately larger than $\ell_0$, not smaller.  To justify in detail the treatment of section~\ref{COLLIDE}, we must {\it assume} that we can arrange $a \ll \ell_0$; then among other nice properties, the motion of vortices is uniformly less than $c_s$, provided only
 \eqn{rLimit}{
  r \gg a \log {\ell_0 \over a} \,.
 }
This last estimate follows from combining \eno{vRatio} with the requirement $v \ll c_s$.

\section{Discussion}
\label{DISCUSSION}

The action \eno{Sall} is an efficient description of interacting vortices because fluctuations in the two-form gauge field that mediate the interactions have already been integrated out.  The bilocal interaction term in \eno{Sall} is complicated for general vortex configurations, but for the specific cases of a single vortex and two identical vortices colliding, this bilocal term leads to the impressive series of elliptic integrals which supplied us with many exact or nearly exact results, such as \eno{Rdefs} and \eno{Sexpansions}.  From a physical perspective, the necessity of cutting off this term provides a notion of renormalization which we have addressed at an elementary level around \eno{ellZeroDef}.  A more sophisticated effective field theory treatment, including a renormalization group equation for the string tension, can be given, but we will not develop it explicitly here; see however \cite{HNPforthcoming}.  A key qualitative feature is that the running tension vanishes at length scales comparable to the scale $\ell_0$ introduced in \eno{ellZeroDef}.  It is natural to expect the system to do interesting things around this dynamically generated length scale.  Indeed, \eno{LambdaSatisfies} indicates that the single vortex instability occurs at wavelengths only moderately larger than $\ell_0$; and according to \eno{UnperturbedEoms}, retrograde motion of small circular vortex loops occurs for $r < \ell_0$.

In our treatment, we have emphasized that the UV cutoff $a$ and the dynamically generated length scale $\ell_0$ are independent quantities, and one may arrange to have $\ell_0 \gg a$ if the rescaled tension $\eta_{\rm bare}$ is negative.  There are reasons to deprecate such an approach.  First: several microscopic models of vortex rings were reviewed in \cite{Donnelly09}, and in all cases $\ell_0 < a$ by a factor of at least $4$ (see Table~1 of \cite{Donnelly09}).  Second: The classical energy \eno{EandP} is negative for $r < \ell_0 e$, so unless $a \gsim \ell_0 e$ the classical theory has a stability problem.  Observing that the quantum zero point energy is positive, one might hold out the hope that quantum effects rescue stability even if $a < \ell_0$; but if this happens, then quantum effects are very important near $\ell_0$ and we would have to ask what justifies use of the classical theory to describe fluctuations around this scale.  Third: Experimental results on single vortices \cite{Widnall73} suggest that finite core size is important in a quantitatively accurate description of the instabilities, and descriptions of colliding vortices often assume significant deformation of the core, as reviewed for example in \cite{Lim95}.  Altogether, real-world applications of vortex ring theory seem to call for a departure from the $\ell_0 \gg a$ regime.  We should however point out that retrograde motion is a remarkable and important feature of rotons in superfluid helium \cite{Tucker19021999}, which is a feature of unperturbed single vortex solutions \eno{UnperturbedEoms} for $r<\ell_0$.  Thus we should not lightly abandon the possibility that dynamics of vortex rings at length scales ${} \lsim \ell_0$ has physical interest.

In any case, we find $\ell_0 \gg a$ a useful starting point because it vastly simplifies the analysis, especially in the case of colliding vortices, while retaining many of the key features, in particular the one and two vortex instabilities.  Moreover, in the $\ell_0 \gg a$ regime, these instabilities are amenable to nearly analytical treatment.  Indeed, the frequencies \eno{omegaE} of single vortex fluctuations and the more complex account in section~\ref{COLLIDE} of colliding vortices are the main technical results of this paper.  In the case of colliding vortices, we saw that the most unstable modes have wavelengths parametrically larger than $\ell_0$ at late times.  So our results in this direction are not wholly dependent on having $\ell_0 \gg a$.

The experimental finding \cite{Lim:1992fk} that the collision of two identical vortex rings at high Reynolds number results in production of many small vortex rings has been a touchstone of the field.  Analytical treatments of an unperturbed head-on collision, resulting only in radial expansion of the colliding rings, date back as far as \cite{Dyson1893}.  As far as we know, ours is the first detailed analytical study that accounts at least qualitatively for the production of many vortices, provided the initial vortices are much larger in diameter than $\ell_0$.  Because our study is based on perturbations around the two original colliding vortices, we can only give an account of the first stages of growth of fluctuations in a mode with large $m$.  The crucial physics of the later stage is reconnection: The vortex lines must come into contact and effect a change of topology.  As far as we can see, this process must be considered in the microscopic theory, and the best we might expect in an effective theory is to parametrize its timescale and its coupling to sound modes.

It may reasonably be asked why experiments such as \cite{Lim:1992fk} on normal fluids at high Reynolds number should be compared with a vortex ring setup derived, as we have done in section~\ref{GROSS} from an effective Gross-Pitaevskii description of a superfluid.  This has been addressed in \cite{Barenghi08,Donnelly09}, where it is argued that well outside the vortex core, where the superfluid density is nearly constant and gradients of it are small, the classical Euler equation applies.

Finally, let us point out a close analogy between vortex rings and giant gravitons.  The latter subject is part of a theme emphasized in \cite{Bigatti:1999iz,Myers:1999ps,McGreevy:2000cw} and related works: in non-commutative spaces, characterized by non-zero form fields, the size of strings, or branes, or brane bound states increases with momentum.  Consider for example 2-branes in ${\bf R}^{4,1}$ in the presence of spatial four-form field strength:
 \eqn{Gfour}{
  G_4 = dC_3 = b \, dx^1 \wedge dx^2 \wedge dx^3 \wedge dx^4 \,.
 }
To make contact with previous results, let us write $x^4 = z$ and choose a gauge such that
 \eqn{CthreeGauge}{
  C_3 = b \, \omega_2 \wedge dz \qquad\hbox{where}\qquad
    d\omega_2 = dx^1 \wedge dx^2 \wedge dx^3 \,.
 }
Consider now a 2-brane in the shape of an $S^2$ in the $x^1$-$x^2$-$x^3$ plane, propagating along the $z$ axis.  A stationary configuration can be found by assuming that only $z$ depends on time, so that the standard action for a two-brane,
 \eqn{TwoBrane}{
  S = -\tau_2 \int d^3 \xi \, \sqrt{-\det g_{\alpha\beta}} + 
    \mu_1 \int C_3 \,,
 }
can be written as $S = \int dt \, L$ where
 \eqn{TwoBraneL}{
  L = -4\pi \tau_2 r^2 \sqrt{1-\dot{z}^2} + {4\pi \over 3} \mu_2 b r^3 \dot{z} \,,
 }
and we are setting the speed of light $c$ equal to $1$.  Evidently, the lagrangian \eno{TwoBraneL} is in close analogy to the free, unperturbed vortex ring lagrangian $L_0$ appearing in the first line of \eno{Lfree}: the only substantive difference is the inertial $\sqrt{1-\dot{z}^2}$ term in \eno{TwoBraneL}.  Furthermore, the 2-brane construction we have just discussed can be embedded in the $AdS_7 \times S^4$ background of M-theory.  This was discussed in detail in \cite{McGreevy:2000cw}; to connect \eno{TwoBraneL} to their work, we need only make the identifications
 \eqn{GiantId}{
  \dot{z} = R \dot\phi \qquad\qquad \mu_2 b = B \,.
 }
Then in the limit $R \to \infty$ with $\dot{z}$ and $r$ held fixed, the lagrangian ${\cal L}_K + {\cal L}_B$ from (3.14) and (3.18) of \cite{McGreevy:2000cw} precisely matches \eno{TwoBraneL}.  Physically, we are permitting motions only on the $S^4$ part of the geometry; we are keeping only the time direction out of $AdS_7$ in the description of the classical motion; and we are keeping the size of the giant graviton fixed while taking the flat space limit in which the number of flux quanta supporting $AdS_7 \times S^4$ is large.  Giant gravitons in $AdS_7 \times S^4$ are stable because they are BPS.  Stability of classical spherical membranes following the dynamics of \eno{TwoBrane} in flat ${\bf R}^{4,1}$ seems closely related, but as we have seen for a single vortex ring, stability depends on the precise properties of the construction in question.

It would be interesting to develop a similar embedding of \eno{Lfree} into string theory in a flat space limit of a background with non-zero $H_3$, for example the NS5-brane.  However, we do not at present understand how to handle the linear dilaton factor in this geometry.  It would also be interesting to inquire how far one can get in describing fluctuations of M2-branes in a limit where second time derivatives are neglected.  Generally speaking, quantum fluctuations of M2-branes are a thorny problem, but perhaps progress can be made in some novel non-relativistic limit.

\section*{Acknowledgments}

We are particularly grateful to B.~Horn, A.~Nicolis, and R.~Penco for stimulating discussions and for insight on the effective field theory aspects of vortex dynamics.  This work was supported in part by the Department of Energy under Grant No.~DE-FG02-91ER40671.

\clearpage
\appendix
\section{Comparison with a vortex filament calculation}
\label{COMPARE}

The stability analysis for a single vortex presented in section~\ref{SINGLE} reproduces the results from \cite{Widnall73} to great analytical and numerical accuracy. Table \eno{TransTable} presents the dictionary for translating between our work and that of \cite{Widnall73}.

\begin{table}[htb]
\begin{center}
\begin{tabular}[b]{|l|c|c|c|c|c|c|}
\hline
 \rule[-7pt]{0pt}{20pt}
& Widnall et al. '73 \cite{Widnall73} & Present work \\ 
\hline
\hline  {\White\Big(} Vortex core radius\footnotemark & $a$ &  $8\,\ell_0\,e^{A-1/2}$\\ 
\hline  {\White\Big(} Vortex ring unperturbed radius & $R$ & $r$ \\ 
\hline  {\White\Big(} UV-cutoff\footnotemark & $l = {a\over 2}e^{1/2-A}$ & $a = {8 \ell_0 \over e} e^{{\eta_{\rm bare} \over \tilde \lambda}}$ \\
\hline  {\White\Big(} Vortex ring axial velocity & $V_0$ & $\dot{z}$\\ 
\hline  {\White\Big(} Amplification rate & $\alpha$ & $i\,\omega_m$ \\ 
\hline  {\White\Big(} Radial perturbation & $\rho_0$ & $r_m$ \\ 
\hline  {\White\Big(} Axial perturbation & $\xi_0$ & $z_m$ \\ 
\hline  {\White\Big(} Vortex strength & $\Gamma$ & $-4\pi \tilde\lambda$\; \\ 
\hline  {\White\Big(} Linearized equations & $\dot{\rho}_0 = V_{\xi_0} \xi_0\,;\; \dot{\xi}_0 = V_{\rho_0} \rho_0$ & equations (\ref{LinearizedEoms}) \\ 
\hline
\end{tabular}
\caption{Formal dictionary for translating between \cite{Widnall73} and present analysis.}
\label{TransTable}
\end{center}
\end{table}
\footnotetext[4]{This, more precisely $a e^{1/2-A} \leftrightarrow 8\ell_0$, is a formal translation between \cite{Widnall73} and the present work. The vertex core radius in the present work is in fact of the order of the UV-cutoff $a$. }
\footnotetext[5]{The r.h.s in both cells describes the relation between the different length scales in the respective papers, but is not part of the dictionary.}

\begin{figure}
  \centerline{\includegraphics[width=5in]{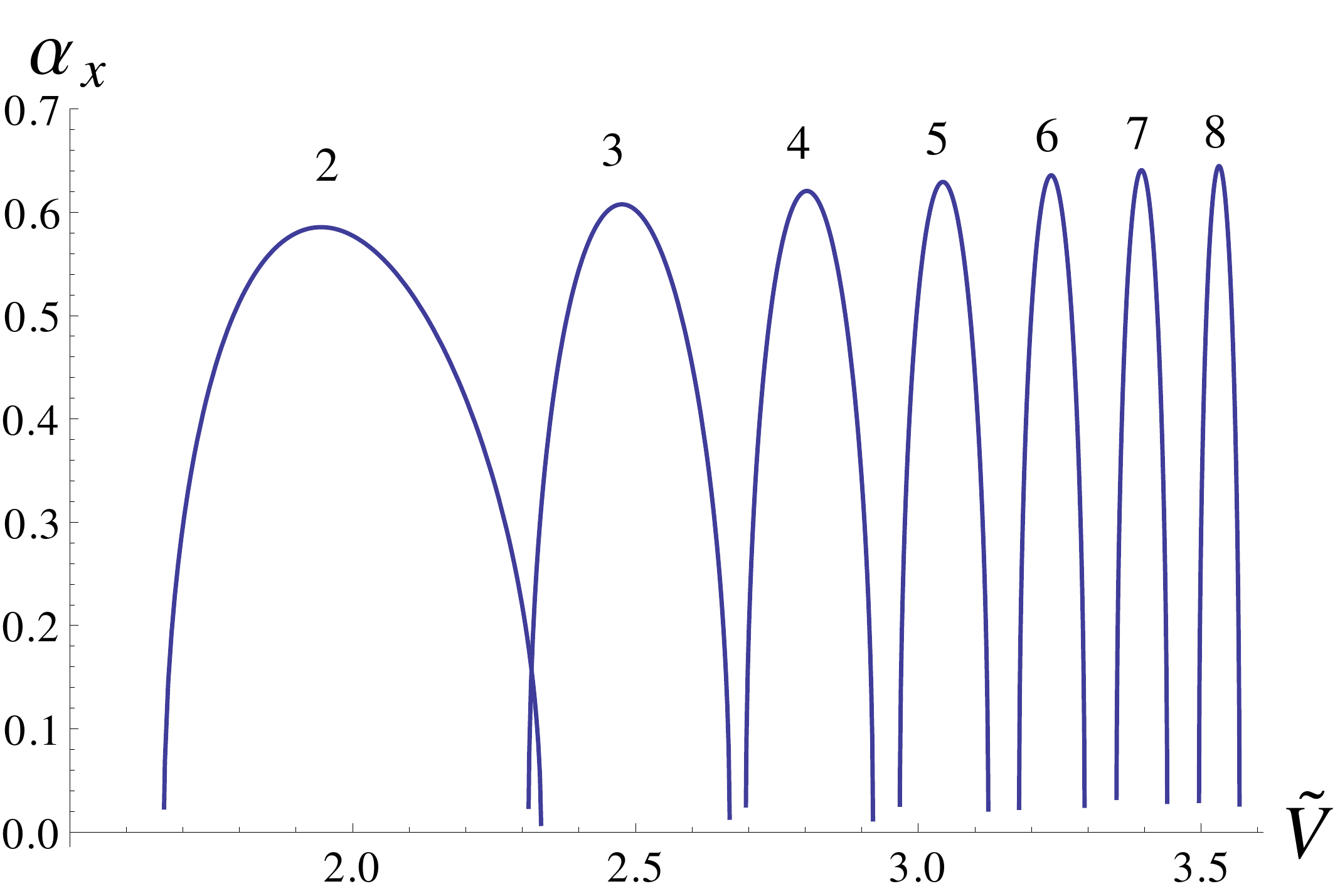}}
  \caption{$\alpha_x$, a measure of vortex ring instability as a function of dimensionless axial velocity $\tilde{V}$ and mode number $m$ (with $\ell_0=1$)}\label{alphaxVtilde}
 \end{figure}

Figure~3 in \cite{Widnall73} plots the ``non-dimensional spatial amplification rate'' $\alpha_x$ as a function of another dimensionless quantity $\tilde{V}$, which is proportional to the axial velocity of the unperturbed vortex, where these quantities are defined to be
\eqn{}{
\alpha_x(\tilde{V};m) = {\alpha R\over V_0}\,, \quad\quad \tilde{V} = {V_0 \over \left(\Gamma/4\pi R\right)}\,.
}
Using the dictionary, one finds
\eqn{alphaxVtildeTrans}{
  \alpha_x \leftrightarrow {i \omega_m r\over \dot{z}}\,,\quad\quad \tilde{V} \leftrightarrow -{\dot{z} r \over \tilde \lambda}\,.
}
On plotting $\alpha_x$ versus $\tilde{V}$ after substituting  for the velocity and exact mode frequencies in (\ref{alphaxVtildeTrans}) by their expressions in (\ref{UnperturbedEoms}) and (\ref{omegaE}), we observe our result: Figure~\ref{alphaxVtilde} matches the ``constant-core-radius'' model of \cite{Widnall73} exceedingly well, with the exception that in \cite{Widnall73} an overlap between the $m=2$ and $m=3$ instabilities was not seen.

As another example, at large $m$ the analogs of \eno{LinearizedEoms}, corresponding to expressions for $V_{\rho_0}$ and $V_{\xi_0}$, were derived in \cite{Widnall73} to be
\eqn{VsWidnall}{
{V_{\rho_0} \over \left(\Gamma/4\pi R^2\right)} &\simeq (m^2-1)\left(\log{2R\over a}-\gamma+1+A\right)-m^2\left(\log m+1\right) +{1\over 4}\log m\,, \cr
{V_{\xi_0} \over \left(\Gamma/4\pi R^2\right)} &\simeq -m^2\left(\log{2R\over a}-\gamma+1+A\right)+\left(m^2-1\right)\left(\log m+1\right) +{1\over 4}\log m\,.
}
In our analysis, the corresponding quantities, $V_{r_m}$ and $V_{z_m}$ are known exactly for any $m$, and can be read off of (\ref{LinearizedEoms}). At large $m$, they become
\eqn{Vs}{
{V_{r_m} \over \left(-\tilde\lambda/ r^2\right)} &\simeq (m^2-1)\left(\log{2 r\over 8 \ell_0 	e^{A-1/2}}-\gamma+1+A\right)-m^2\left(\log m+1\right) +{1\over 4}\log m \cr 
	& \qquad +{59\over 24}-{3\over 2}\log 2-{3\over 4}\gamma + O\left({1\over m^2}\right)\,, \cr
{V_{z_m} \over \left(-\tilde\lambda/ r^2\right)} &\simeq -m^2\left(\log{2 r\over 8 \ell_0 e^{A-1/2}}-\gamma+1+A\right)+\left(m^2-1\right)\left(\log m+1\right) +{1\over 4}\log m \cr
	& \qquad +{25\over 24}-{3\over 2}\log 2-{3\over 4}\gamma + O\left({1\over m^2}\right)\,,
} 
where we have written $V_{r_m}$ and $V_{z_m}$ in such a manner so as to make the dictionary between the hydrodynamics treatment \cite{Widnall73} and our calculation explicit. The results match up to $O(1)$ discrepancies.

\begin{figure}
  \centerline{\includegraphics[width=5in]{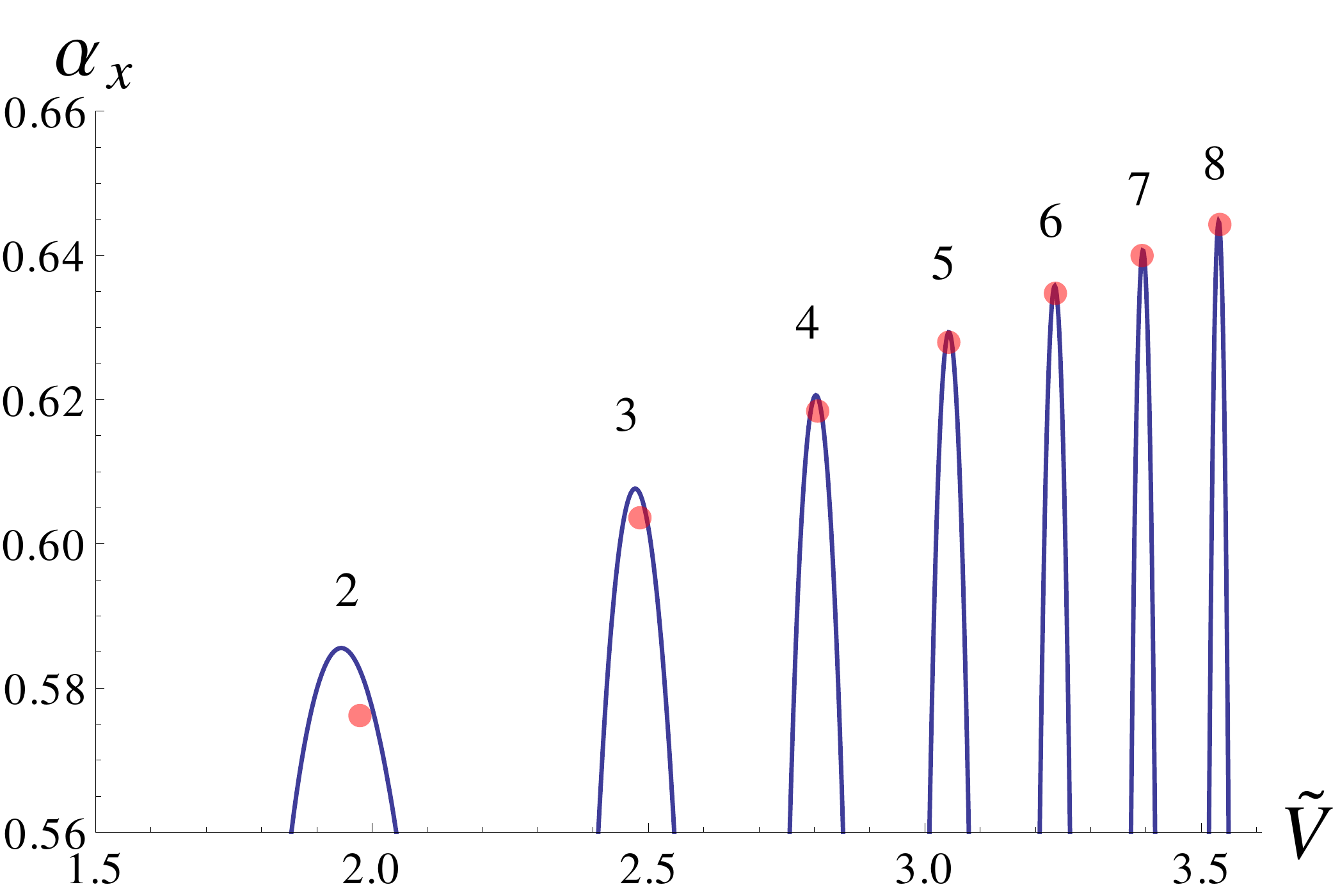}}
  \caption{(Color online.)  Exact computation for the amplification rate (solid lines), and large $m$ approximation (\ref{alphax}) for the peaks (red points).}\label{alphaxVtildePeaks}
 \end{figure}

We will now derive analytical expressions for the ``instability band'' and the ``maximum amplification rate'' as a function of large mode number $m$, closely following the calculation in \cite{Widnall73}. In figure \ref{alphaxVtilde}, the ``instability band'' corresponds to the lower and upper limits on $\tilde V$ which yield an instability for mode $m$, while the ``maximum amplification rate'' corresponds to the maximum $\alpha_x$ at that mode $m$, corresponding to maximum instability. As can be seen in figure \ref{alphaxVtildePeaks} which shows the result of this computation, the large $m$ approximation works well even at moderate values of $m$ (such as 8). The main difference between our computation and that in \cite{Widnall73} is that we keep $O(1)$ terms in our analysis. The calculation proceeds as follows.

The upper limit for the instability band is obtained by setting $V_{z_m}=0$ in (\ref{Vs}), solving for $\ell_0/r$, and substituting the result into (\ref{alphaxVtildeTrans}) to obtain $\tilde V$: 
\eqn{UpperVtilde}{
\log {\ell_0 \over r} &= { -\left(m^2-1\right)\left(\log m+1\right)-{1\over 4}\log m+ m^2 \left({3 \over 2}-\gamma -\log 4\right)-\frac{25}{24}+{3 \over 4}\gamma+{3 \over 2} \log 2 \over m^2} + O\left({1 \over m^4}\right)\,, \cr
\tilde V &= \log{r\over\ell_0} = \left(1-{3\over 4m^2}\right)\log m + 2\log 2-{1\over 2}+\gamma+{1\over m^2}\left({1\over 24}-{3\over 4}\gamma-{3\over 2}\log 2\right) + O\left({1 \over m^4}\right)\,.
}
The lower limit for the instability band is obtained by setting $V_{r_m}$ in (\ref{Vs}), expanded about $\ell_0/r + \delta$ for small $\delta$, to zero and solving for $\delta$: 
\eqn{deltal0}{
\delta &= -{3\over 8m^3}\left(\log m -1+\gamma+2\log 2\right)e^{1/2-\gamma}+O\left({\log m\over m^5}\right)\,, \cr
{\delta \over \ell_0/r} &= -{3\over 2}\; {\log m-1+\gamma+2\log 2 \over m^2}+O\left({\log m\over m^4}\right)\,.
}
Maximum amplification occurs approximately in the middle of the band, corresponding to 
\eqn{l0rMax}{
\left({\ell_0\over r}\right)_{\rm max}={\ell_0\over r} + {\delta\over  2} &\simeq {\ell_0 \over r}\left(1-{3\over 4}\; {\log m-1+\gamma+2\log 2\over m^2}\right)\,,
}
where $\ell_0/r$ is given by (\ref{UpperVtilde}). The corresponding amplification rate $\alpha_x$ can be computed by substituting (\ref{l0rMax}) into (\ref{alphaxVtildeTrans}), and expanding to $O(1)$ at large $m$, 
\eqn{alphax}{
\left(\alpha_x\right)_{\rm max} \simeq {3\over 4}\;{\log m +\gamma -1+2 \log 2 \over \log m+ \gamma -{1\over 2}+2\log 2} \approx {3\over 4}\;{\log m +\gamma -1+2 \log 2 \over\tilde V_{\rm max}} \approx  {3\over 4}\;{\log m +0.96351 \over\tilde V_{\rm max}}\,,
}
where
\eqn{VMax}{
\tilde V_{max} \approx \log {m}+\gamma-{1\over 2} +2\log 2-{17\over 24 m^2}\,.
}
In contrast, the authors in \cite{Widnall73} obtained $\alpha_x \simeq 3/4$ which is accurate only at very large $m$, but noted that a better fit to their numerical results was given by
\eqn{alphaxWidnall}{
\alpha_x \simeq {3\over 4}\;{\log m +1 \over\tilde V}\,,
}
which closely matches our result (\ref{alphax}) obtained analytically. In figure \ref{alphaxVtildePeaks} we superimposed $\alpha_x$ as obtained in (\ref{alphax}) and $\tilde V$ at the radius of maximum instability (\ref{l0rMax}), as quoted in \eno{VMax}, on the exact result shown in figure \ref{alphaxVtilde}. As can be seen, the large $m$ approximation works very well even for $m$ as low as $8$.

\bibliographystyle{ssg}
\bibliography{vortex}
\end{document}